\documentclass[12pt]{article}
\usepackage{amsmath,amsthm,amscd,amsfonts,amssymb}
\usepackage{latexsym}
\usepackage[usenames]{color}

\newcommand{\BB}{\mathcal B}

\newcommand{\RR}{\mathbb R}

\newcommand{\ZZ}{\mathbb Z}
\newcommand{\EE}{\mathbb E}
\newcommand{\NN}{\mathbb N}

\newcommand{\hh}{\mathcal H}

\newcommand{\nn}{\mathcal N}

\newtheorem{thm}{Theorem}[section]

\newtheorem{cor}[thm]{Corollary}

\newtheorem{rem}[thm]{Remark}

\author{Werner Kirsch \\ Fakult\"at Mathematik \\ Fern Universit\"at Hagen \\
Hagen , Germany \\ email:werner.kirsch@fernuni-hagen.de \\
M Krishna \\ Ashoka University \\Plot No 2, Rajiv Gandhi Education City \\ Rai, Haryana 131029, India\\ email: krishna.maddaly@ashoka.edu.in}
\title{Analyticity of density of states for the Cauchy distribution}
\begin{document}
\maketitle
\begin{abstract}
We compute the density of states for the Cauchy distribution for a large class of random operators and show it is analytic in a strip about the real axis.
\end{abstract}

\section{The Models and Results}

In this short note we show that many random operators with the 
Cauchy distribution have analytic density of states, independent
of disorder.  This theorem applied to the Anderson model
shows analyticity through the 
mobility edge (or the mobility band)  on the Bethe lattice
for small disorder and some models of Random Schr\"odinger
operators.

\newpage
The density of states of random operators is of interest for many
random models.  We are concerned in particular with random models
coming from i.i.d random variables with the Cauchy distribution.  For these
models there is an exact formula for the density of states.

One of the earliest models with the Cauchy distribution was the Lloyd model
studied by Lloyd \cite{Lloyd} on $\ell^2(\ZZ^3)$, in which he obtained
an exact formula for the density of states.  In the book of Carmona-Lacroix
\cite{CL}, page 329 and in problem VI.5.5 one finds calculations
giving the exact value for the integrated density of states (IDS) 
for the Anderson model on the lattice $\ell^2(\ZZ^d)$ with the Cauchy
distribution.  On the Bethe lattice Accosta-Klein \cite{MR1184777}, proved
that the density of states is analytic in a strip about the real axis
for any distribution close to and including the Cauchy distribution,
in a function space.  We note that Accosta-Klein \cite{MR1184777} result
is the first known result where smoothness of DOS is shown in the 
region where absolutely continuous spectrum is present for random
operators.    

In this note we show that the exact expressions for the IDS
are valid in a much larger context both for the discrete and some continuous
models, whenever i.i.d random variables with the Cauchy distribution
are involved.  We use the Trotter product formula for proving our result.  

We consider a separable Hilbert space $\hh$
and a self-adjoint operator $H_0$ 
with domain $D(H_0) \subset \hh$, a sequence $\{P_n\}$ of 
mutually orthogonal projections
such that $\sum_{n\in \NN} P_n = Id$  and consider the
operator $\displaystyle{N^\alpha = \sum_{n \in \NN} n^\alpha P_n , ~~ \alpha > 1,}$ 
with domain $D(N^\alpha) \subset \hh$.  We consider i.i.d random variables 
$\{\omega_n, ~ n \in \NN\}$ with the Cauchy distribution whose density is given by 
$\psi_\lambda(x) = \frac{1}{\pi}\frac{\lambda}{\lambda^2+ x^2} , ~
\lambda >0$.  If we consider an $\alpha>1$, %$0 < \alpha <1$, 
then under the 
assumption on the random variables $\{\omega_n\}$,
an application of the Borel-Cantelli Lemma implies that the set 
\begin{align}\label{omegazero}
\Omega_0 = \{(\omega_n) : |\omega_n| \leq |n|^\alpha, ~ \mathrm{for ~ all ~ but ~ finitely ~ many} ~ n \}
\end{align}
has probability $1$.  Therefore  the domain of the self-adjoint operator 
$V^\omega = \sum_{n\in \NN} \omega_nP_n$ contains  $D(N^\alpha)$
for every $\omega \in \Omega_0$.  We assume that 
$D(H_0) \cap D(N^\alpha)$ is dense in $\hh$ and that the random operators
\begin{align}\label{themodel}
H^\omega = H_0 + V^\omega, ~~ V^\omega = \sum_{n\in \NN} \omega_n P_n  
\end{align}
are essentially self-adjoint on 
$D(H_0) \cap D(N^\alpha)$ for every $\omega \in \Omega_0$.
In the following we denote by $E_{A}()$, the spectral projection of
a self-adjoint operator $A$.

Then we have, with $\EE$ denotes integration  with respect to $\omega$   
and $*$ denotes convolution :
\begin{thm}\label{Cauchy}
Let $\hh$ be a separable Hilbert space and $H_0, V^\omega$ be self-adjoint
operators on $\hh$, as in equation (\ref{themodel}), with 
$D(V^\omega) \supset {\mathcal D}$ for some
dense set ${\mathcal D}$ for almost every $\omega$. Assume that  
$D(H_0) \cap {\mathcal D}$
is dense in $\hh$ with $H^\omega$ essentially
self-adjoint on $D(H_0) \cap {\mathcal D}$ for allmost all $\omega$. 
Let $\phi, \psi \in \hh$ be unit vectors and set
$\mu_{\phi,\psi} = \langle \phi, E_{H_0}() \psi\rangle$.
Then,
\begin{align}\label{IDS}
\nn(\cdot) = \EE \bigg(\langle \phi, E_{H^\omega}(\cdot) \psi\rangle \bigg)
= \psi_\lambda * \mu_{\phi,\psi}, 
\end{align}
as measures. 
\end{thm}

We have a few corollaries of this theorem applied to the Anderson model
on $\ZZ^d$ and the Bethe lattice, where we denote by $\delta_0$
the unit vector supported at $0$ in $\ZZ^d$ or a chosen and fixed root
in the Bethe lattice.  In both these cases, the operator $H_0$, being the
adjascency matrix, is bounded.  Therefore the conditions of the 
Theoremm \ref{Cauchy} are satisfied and the following corollary results
by setting $\phi =\psi = \delta_0$ in the Theorem \ref{Cauchy}. 

\begin{cor}\label{corollary1}
Consider the Anderson model on $\ell^2(\ZZ^d)$ or $\ell^2(\BB)$,
$\BB$ the Bethe lattice, with
the single site distribution $\psi_\lambda(x)dx
, ~ \lambda >0$.  Then the density of states, which is the density of the 
measure 
$$
\nn(\cdot) = \EE\big(\langle \delta_0, E_{H^\omega}(\cdot) \delta_0\rangle\big),
$$
is analytic in a strip $\{z : \Im(z) < \lambda\}$ about the real axis.
\end{cor} 

\begin{rem}\label{remark1}
The above Corollary was proved for the three dimensional case by
Lloyd \cite{Lloyd} and a proof is indicated in the problem VI.5.5 
of Carmona \cite{CL} in the case of the 
Anderson model on $\ell^2(\ZZ^d)$, for any disorder $\lambda >0$.
Our result as that of Accosta-Klein \cite{MR1184777}, shows smoothness
of DOS through the mobility edge and everywhere in the spectrum
, when applied to the Bethe lattice where existence of absolutely 
continuous spectrum is proved see Klein \cite{K1}, Froese et.al. \cite{FHS}
and Aizenman-Warzel \cite{AW}.  Our results may also apply to the models considered
by Anantharaman et.al. \cite{AISW}.
\end{rem}

The second corollary is for Random Schr\"odinger Operators on $L^2(\RR)$.  
Consider a collection of smooth bump functions $\{u_n, n \in \ZZ\}$ supported 
in a neighbourhood of unit cubes centered at $n \in \ZZ$. Let,
\begin{align}\label{continuous}
H^\omega = -\Delta + V^\omega, ~~ V^\omega(x) = \sum_{n \in \ZZ} \omega_n u_n(x), ~~ \sum_{n} u_n(x) = 1, ~~ \forall  x \in \RR,
\end{align}
where $-\Delta$ is the Laplacian defined on its maximal domain $D(-\Delta)$,
 $\{\omega_n\}$  
are i.i.d random variables with Cauchy distribution $\psi_\lambda(x) dx$.

Then we have the corollary, where as before $*$ denotes
convolution.  The only statement to verify to prove the
corrolary is the essential self-adjointness of the random Schr\"odinger
operators involved, which follows from a theorem of  
Kirsch-Martinelli \cite{MR726328}.  We note that if the essential
self-adjointness holds for the operators on $L^2(\RR^d)$, then the
same result extends to that case also.

\begin{cor}\label{corcts}
Consider the random Schr\"odinger operators given in equation 
(\ref{continuous}).  Then there exists a set $\Omega_0$ of probability 1
such that for each $\omega \in \Omega_0$, the operators $H^\omega$
are essentialy self-adjoint on $D(-\Delta) \cap D(|x|^\alpha)$, for some
$\alpha > 1$ and the integrated density of states $\nn$ is given by
\begin{align}\label{idscts}
\nn(E) = \frac{ \psi_\lambda * tr\bigg( u_0 E_{H_0}((-\infty, E]))\bigg)}{\int u_0(x) dx}, ~~ E \in \RR.
\end{align}

\end{cor}

\section{The Proofs}

\vspace{.5cm}

{\noindent \bf Proof of Theorem \ref{Cauchy}:} We first note that for any
unit vector $\phi \in \hh$, $\mu_\phi$ is a probability measure and 
its convolution with $\psi_\lambda$ is also a probability measure.

The proof of this is by the use of Trotter product formula. 

We start by choosing a sequence $Q_M$ of finite rank 
orthogonal projections such that
\begin{align}\label{theqms}
&\lim_{M\rightarrow \infty} Q_M = Id, ~~ [Q_M , P_n] = 0, ~~  Q_{M,n} = Q_M P_n  ~~ \forall ~ n \in \NN \\
&\forall ~ M \exists ~ N(M) ~ s.t. ~ Rank(Q_{M,n}) \neq 0, \forall ~  n \leq N(M), ~ N(M) \rightarrow \infty ~ \mathrm{as} ~ M \rightarrow \infty,  
\end{align}
where the limits are in the stong sense.  The existence of such $Q_M$ is not
hard to see, since $P_n$s are mutually orthogonal and add to the identity.

We will prove the theorem by showing that the Fourier transform of
$\nn$ is the product of Fourier transforms of $\mu_{\phi, \psi}$ and $\psi_\lambda$.    
By the spectral theorem for self-adjoint operators, the Fourier transform 
of $\nn$ is given by 
\begin{align}\label{eq0}
\int e^{itx} ~ d\nn(x) = \EE\big( \langle \phi, e^{itH^\omega} \psi\rangle\big).
\end{align}
We set $V^\omega = \sum\omega_n P_n$ and use the Trotter product formula
%\cite[Theorem VIII.30]{Simon1} and its stronger version 
\cite[Theorem VIII.31]{RS1}, to write the right hand side of equation (\ref{eq0}) as
\begin{align}\label{eq1}
&\EE\big( \langle \phi, e^{itH^\omega} \psi\rangle\big) = \lim_{k\rightarrow \infty} \EE\big( \langle \phi, \big(e^{i\frac{t}{k}H_0} e^{i\frac{t}{k}V^\omega}\big)^k \psi\big) \nonumber \\
&= \lim_{k\rightarrow \infty}\lim_{M\rightarrow \infty} \EE\big(\langle\phi, \big( e^{i\frac{t}{k}H_0} e^{i\frac{t}{k}V^\omega} Q_{M}\big)^k   \psi\rangle \big)\nonumber \\
&= \lim_{k\rightarrow \infty}\lim_{M\rightarrow \infty} \sum_{n_1, \dots, n_k =1}^{N(M)}\EE\bigg( \langle\phi, \big(\prod_{j=1}^{\substack{k\\\rightarrow}} e^{i\frac{t}{k}H_0} e^{i\frac{t}{k}V^\omega} Q_{M,n_j}\big)   \psi\rangle \bigg),
\end{align} 
where in the second equality we used the Lebesgue dominated convergence theorem
to interchange the limit and expectation and in the next equality the arrow
on the product denotes an ordered product with increasing index $j$ and finally
the sum and the expectation are interchanged
using Fubini's theorem.  Since we have from definitions of
$V^\omega, Q_{M,n}$ that $e^{i\frac{t}{k} V^\omega} Q_{M, n_j} = e^{i\frac{t}{k} \omega_{n_j}}$,
 the above equality becomes 
\begin{align}\label{eq2}
&\EE\big( \langle \phi, e^{itH^\omega} \psi\rangle\big)\nonumber \\
&= \lim_{k\rightarrow \infty}\lim_{M\rightarrow \infty} \sum_{n_1, \dots, n_k =1}^{N(M)}\EE\bigg( \langle\phi, \big(\prod_{j=1}^{\substack{k\\\rightarrow}} e^{i\frac{t}{k}H_0} e^{i\frac{t}{k}\omega_{n_j}} Q_{M,n_j}\big)   \psi\rangle \bigg)
\nonumber \\
&= \lim_{k\rightarrow \infty}\lim_{M\rightarrow \infty} \sum_{n_1, \dots, n_k =1}^{N(M)}e^{-\lambda |t|}  \langle\phi, \big(\prod_{j=1}^{\substack{k\\\rightarrow}} e^{i\frac{t}{k}H_0} Q_{M,n_j}\big)   \psi\rangle \bigg)\nonumber \\
&= \lim_{k\rightarrow \infty}\lim_{M\rightarrow \infty} e^{-\lambda |t|}  \langle\phi, \big(\prod_{j=1}^{\substack{k\\\rightarrow}} e^{i\frac{t}{k}H_0} Q_{M}\big)   \psi\rangle \bigg) \nonumber \\
&= e^{-\lambda |t|} \langle \phi, e^{itH^\omega} \psi\rangle,
\end{align}
where in the second equality we used the fact that when the
product is expanded for a fixed configuration of $(n_1, \dots, n_k)$,
we will get an expression of the form
$$
e^{i\frac{t}{k} (\alpha_1 \omega_{n_1} + \alpha_2 \omega_{n_2} + \dots + \alpha_k \omega_{n_k})},
$$
with the $\alpha_j$'s denoting the number of times the $\omega_{n_j}$ occurs
in the product and for any configuration of $(n_1, \dots, n_k)$ we have
$\sum \alpha_j = k$.  This fact together with 
the fact that $\omega_{n_j}$s are
independent random variables for distint $n_j$ and the expectation with
respect to the Cauchy distribution satisfies
$$
\int e^{isx} \psi_\lambda(x) dx = e^{-\lambda |s|}, ~~ \EE e^{is \omega_n} e^{iw\omega_m}  = e^{-\lambda |s+w|}, ~ n\neq m 
$$
shows that the expectation always gives the value $e^{-\lambda |t|}$ for
any configuration of $(n_1, \dots, n_k)$ in the sum, so we could take
that factor out and resum the expression.  

The equation (\ref{eq2}) shows that $\EE\langle \phi, e^{itH^\omega} \psi\rangle$
is integrable as a function of $t$ and so by Fourier inversion, the
Theorem follows. \qed

\vspace{.5cm}

{\noindent \bf Acknowledgement:}  We thank K B Sinha and Peter Hislop for conversations on the Trotter product formula.

\bibliographystyle{plain}
\bibliography{Cauchy}

\begin{thebibliography}{1}

\bibitem{MR1184777}
Victor Acosta and Abel Klein.
\newblock Analyticity of the density of states in the {A}nderson model on the
  {B}ethe lattice.
\newblock {\em J. Statist. Phys.}, 69(1-2):277--305, 1992.

\bibitem{AW}
Michael Aizenman and Simone Warzel.
\newblock Resonant delocalization for random {S}chr\"{o}dinger operators on
  tree graphs.
\newblock {\em J. Eur. Math. Soc. (JEMS)}, 15(4):1167--1222, 2013.

\bibitem{AISW}
Nalini Anantharaman, Maxime Ingremeau, Mostafa Sabri, and Brian Winn.
\newblock Absolutely continuous spectrum for quantum trees.
\newblock {\em arXiv preprint arXiv:2003.12765}, 2020.

\bibitem{CL}
Ren{\'e} Carmona and Jean Lacroix.
\newblock {\em Spectral theory of random Schr{\"o}dinger operators}.
\newblock Springer Science \& Business Media, 2012.

\bibitem{FHS}
Richard Froese, David Hasler, and Wolfgang Spitzer.
\newblock Absolutely continuous spectrum for the anderson model on a tree: a
  geometric proof of klein's theorem.
\newblock {\em Communications in mathematical physics}, 269(1):239--257, 2007.

\bibitem{MR726328}
Werner Kirsch and Fabio Martinelli.
\newblock On the essential selfadjointness of stochastic {S}chr\"{o}dinger
  operators.
\newblock {\em Duke Math. J.}, 50(4):1255--1260, 1983.

\bibitem{K1}
Abel Klein.
\newblock Extended states in the anderson model on the bethe lattice.
\newblock {\em Advances in Mathematics}, 133(1):163--184, 1998.

\bibitem{Lloyd}
P.~Lloyd.
\newblock Exactly solvable model of electronic states in a three dimensional
  disordered hamiltonian: non existence of localized states.
\newblock {\em J. Physics (C)}, 2:1717--1725, 1969.

\bibitem{RS1}
Michael Reed and Barry Simon.
\newblock {\em Methods of modern mathematical physics: Functional analysis},
  volume~1.
\newblock Academic Press Inc., 1980.

\end{thebibliography}

\end{document}